\documentstyle[12pt]{article} \hoffset -0.5in \textwidth 6.50in \textheight
9.00in  \parskip 7pt \openup0.1\jot \parindent=0.5in
\newskip\humongous \humongous=0pt plus 1000pt minus 1000pt
\def\caja{\mathsurround=0pt}
\def\eqalign#1{\,\vcenter{\openup1\jot \caja
        \ialign{\strut \hfil$\displaystyle{##}$&$
        \displaystyle{{}##}$\hfil\crcr#1\crcr}}\,}
\newif\ifdtup

\def\eqright #1\cr{\noalign{\hfill$\displaystyle{{}#1}$}}
\def\eqleft #1\cr{\noalign{\noindent$\displaystyle{{}#1}$\hfill}}

\def\oldreffmt#1{\rlap{[#1]} \hbox to 2\parindent{}}

\def\figfmt#1{\rlap{Figure {#1}} \hbox to 1in{}}

\def\sectioneq{\def\theequation{\thesection.\arabic{equation}}{\let
\holdsection=\section\def\section{\setcounter{equation}{0}\holdsection}}}%

\newcounter{holdequation}

\def\auto{\eqno(\refstepcounter{equation}\theequation)}
\def\begineq #1\endeq{$$ \refstepcounter{equation}\eqalign{#1}\eqno
        (\theequation) $$}
\def\contlimit{\,{\hbox{$\longrightarrow$}\kern-1.8em\lower1ex
\hbox{${\scriptstyle (a\rightarrow0)}$}}\,}
\def\centeron#1#2{{\setbox0=\hbox{#1}\setbox1=\hbox{#2}\ifdim
\wd1>\wd0\kern.5\wd1\kern-.5\wd0\fi
\copy0\kern-.5\wd0\kern-.5\wd1\copy1\ifdim\wd0>\wd1
\kern.5\wd0\kern-.5\wd1\fi}}
\def\centerover#1#2{\centeron{#1}{\setbox0=\hbox{#1}\setbox
1=\hbox{#2}\raise\ht0\hbox{\raise\dp1\hbox{\copy1}}}}
\def\centerunder#1#2{\centeron{#1}{\setbox0=\hbox{#1}\setbox
1=\hbox{#2}\lower\dp0\hbox{\lower\ht1\hbox{\copy1}}}}
\def\lsim{\;\centeron{\raise.35ex\hbox{$<$}}{\lower.65ex\hbox
{$\sim$}}\;}
\def\gsim{\;\centeron{\raise.35ex\hbox{$>$}}{\lower.65ex\hbox
{$\sim$}}\;}
\def\st#1{\centeron{$#1$}{$/$}}

\def\super#1{\ifmmode \hbox{\textsuper{#1}}\else\textsuper{#1}\fi}
\def\textsuper#1{\newcount\holdspacefactor\holdspacefactor=\spacefactor
$^{#1}$\spacefactor=\holdspacefactor}

\def\getcite#1,{\advance\citenumber by1
\def\getcitearg{#1}\def\lastarg{@}
\ifnum\citenumber=1
\ref{#1}\let\next=\getcite\else\ifx\getcitearg\lastarg\let\next=\relax
\else ,\ref{#1}\let\next=\getcite\fi\fi\next}

\def\pom{{\rm P\kern -0.53em\llap I\,}}
\def\spom{{\rm P\kern -0.36em\llap \small I\,}}
\def\sspom{{\rm P\kern -0.33em\llap \footnotesize I\,}}

\relax
\newskip\humongous \humongous=0pt plus 1000pt minus 1000pt
\def\caja{\mathsurround=0pt}
\def\eqalign#1{\,\vcenter{\openup1\jot \caja
        \ialign{\strut \hfil$\displaystyle{##}$&$
        \displaystyle{{}##}$\hfil\crcr#1\crcr}}\,}
\newif\ifdtup

\def\eqright #1\cr{\noalign{\hfill$\displaystyle{{}#1}$}}
\def\eqleft #1\cr{\noalign{\noindent$\displaystyle{{}#1}$\hfill}}

\def\oldreffmt#1{\rlap{[#1]} \hbox to 2\parindent{}}

\def\figfmt#1{\rlap{Figure {#1}} \hbox to 1in{}}

\def\auto{\eqno(\refstepcounter{equation}\theequation)}
\def\begineq #1\endeq{$$ \refstepcounter{equation}\eqalign{#1}\eqno
        (\theequation) $$}
\def\contlimit{\,{\hbox{$\longrightarrow$}\kern-1.8em\lower1ex
\hbox{${\scriptstyle (a\rightarrow0)}$}}\,}
\def\centeron#1#2{{\setbox0=\hbox{#1}\setbox1=\hbox{#2}\ifdim
\wd1>\wd0\kern.5\wd1\kern-.5\wd0\fi
\copy0\kern-.5\wd0\kern-.5\wd1\copy1\ifdim\wd0>\wd1
\kern.5\wd0\kern-.5\wd1\fi}}
\def\centerover#1#2{\centeron{#1}{\setbox0=\hbox{#1}\setbox
1=\hbox{#2}\raise\ht0\hbox{\raise\dp1\hbox{\copy1}}}}
\def\centerunder#1#2{\centeron{#1}{\setbox0=\hbox{#1}\setbox
1=\hbox{#2}\lower\dp0\hbox{\lower\ht1\hbox{\copy1}}}}
\def\lsim{\;\centeron{\raise.35ex\hbox{$<$}}{\lower.65ex\hbox
{$\sim$}}\;}
\def\gsim{\;\centeron{\raise.35ex\hbox{$>$}}{\lower.65ex\hbox
{$\sim$}}\;}
\def\st#1{\centeron{$#1$}{$/$}}

\def\super#1{\ifmmode \hbox{\textsuper{#1}}\else\textsuper{#1}\fi}
\def\textsuper#1{\newcount\holdspacefactor\holdspacefactor=\spacefactor
$^{#1}$\spacefactor=\holdspacefactor}

\def\getcite#1,{\advance\citenumber by1
\ifnum\citenumber=1
\ref{#1}\let\next=\getcite\else\ifx#1@\let\next=\relax
\else ,\ref{#1}\let\next=\getcite\fi\fi\next}

\def\upon #1/#2 {{\textstyle{#1\over #2}}}
\relax

\def\subhead#1{\bigskip\vbox{\noindent\bf #1}\nobreak\par}

\def\til#1{\centeron{\hbox{$#1$}}{\lower 2ex\hbox{$\char'176$}}}
\def\tild#1{\centeron{\hbox{$\,#1$}}{\lower 2.5ex\hbox{$\char'176$}}}
\def\sumtil{\centeron{\hbox{$\displaystyle\sum$}}{\lower
-1.5ex\hbox{$\widetilde{\phantom{xx}}$}}}

\def\pom{{\rm P\kern -0.53em\llap I\,}}
\def\spom{{\rm P\kern -0.36em\llap \small I\,}}
\def\sspom{{\rm P\kern -0.33em\llap \footnotesize I\,}}

\begin{document}
\begin{titlepage}
\rightline{\vbox{\halign{&#\hfil\cr
&ANL-HEP-CP-94-49\cr
&\today\cr}}}

{}~
\vspace{1in}

\begin{center}

{\bf HIGHER-ORDER LIPATOV KERNELS AND THE QCD POMERON}

\medskip

Alan R. White
\footnote{Work supported by the U.S. Department of Energy, Division of High
Energy Physics, Contract\newline W-31-109-ENG-38}
\\ \smallskip
High Energy Physics Division, \\
Argonne National Laboratory, \\
Argonne, IL 60439. \\

\end{center}

\begin{abstract}

Three closely related topics are covered. The derivation of $O(g^4)$ Lipatov
kernels in pure glue QCD. The significance of quarks for the physical
Pomeron in QCD. The possible inter-relation of Pomeron dynamics with
Electroweak
symmetry breaking.

\end{abstract}

\vspace{3in}

\noindent Invited Talk presented at the Workshop on Quantum InfraRed
Physics, Paris, France, June 1994 and the XVII Kazimierz Meeting on Elementary
Particle Physics, Kazimierz, Poland, May 1994.

\end{titlepage}

\subhead{1. Introduction}

I will spend most of my time in this talk describing my recent
derivation\cite{ker} of
higher-order Lipatov kernels, for pure-glue QCD, directly from $t$- channel
unitarity. However, I also want to put this analysis in the context of my
general study\cite {arw2} of the physical, or {\it soft}, Pomeron in full
QCD with quarks. I shall outline why quarks play such an important part in
the emergence of confinement in the low transverse momentum region. I will
also describe how study of the full unitarity properties of the Pomeron
leads naturally to the introduction of a new higher-color quark sector which
can provide a very attractive picture of dynamical electroweak symmetry
breaking.

\subhead{2. Higher-order Lipatov Kernels}

Currently there is much excitement about the small-x behavior of
structure functions. In particular it seems that at HERA the ``Lipatov
Pomeron''\cite{lip} may have been seen i.e.
$$
\eqalign { F_2(x,q^2) ~\sim ~ x^{1-\alpha_0} ~\sim ~x^{-{1 \over 2}}}
\auto
$$
The Lipatov equation was originally derived from extensive
leading and next-to-leading log calculations in the Regge limit of (massive)
Yang-Mills theories\cite{bfkl,cl} and is applied as an evolution equation for
parton distributions at small-x i.e.
$$
\eqalign{ {\partial \over \partial (ln {1 / x})}F(x,q^2) ~=~\tilde{F}(x,q^2)~+~
{1 \over (2\pi)^3}\int {d^2k \over k^4} ~K(k,q) F(x,k^2) }
\auto
$$
$K(k,q)$ is given in terms of the $O(g^2)$ Lipatov kernel by
$$
\eqalign{ K(k,q)~=~K^{(2)}_{2,2}(k,-k,q,-q)}
\auto
$$
where
$$
\eqalign{ {2 \over 3g^2}K^{(2)}_{2,2}(k_1,k_2,k_3,k_4)~=&
{}~\sum_{\scriptscriptstyle 1<->2}~
\Biggl(~(2\pi)^3~k_1^2J_1(k_1^2)k_2^2~\Bigl(k_3^2\delta^2(k_2-k_4)
{}~+~k_4^2\delta^2(k_2-k_3)\Bigr)\cr
&~-~{k_1^2k_4^2~+~k_2^2k_3^2 \over (k_1-k_3)^2}~
-~(k_1+k_2)^2\Biggr)}
\auto\label{2,2}
$$
To obtain non-leading corrections to the $O(g^2)$ kernel, it
appears that (very complicated) very non-leading log {\bf Regge limit
calculations} are required.

Back in the days when Pomeron Reggeon Field Theory (RFT) was studied
intensely, it was understood that to satisfy {\bf multiparticle t-channel
unitarity} in the Regge limit a theory must be describable  in terms of {\bf
reggeon diagrams} that satisfy {\bf reggeon unitarity}\cite{arw1}. For an
even signature Pomeron, reggeon unitarity is very simple - as we now
briefly review.

Introducing RFT variables,~ $E = 1-\ell$ and $k^2=-t$~, and writing
$\alpha_{\spom}(t) = 1 - \Delta(k^2)$, a ``partial-wave amplitude''
$a^+(\ell,t)$ will satisfy reggeon unitarity if we can write
$$
a^+(\ell,t)\equiv F\left(E,k^2\right)=\sum^\infty_{n,m=1}
F_{nm}(E,k^2),
\auto
$$
with
$$
\eqalign{
F_{nm}&\left(E,k^2\right){}=
        {1 \over (2\pi)^{3n+3m}} \int\left[\prod_{i,j} d^2k_id^2k_j
        \delta^2 \left(k-\sum^n_{i=1} k_i\right)\delta^2 \left(k-\sum^m_{j=1}
        k'_j\right)\right]\cr
 &g_n\left(k_1,\ldots k_n\right)g_m\left(k_1',\ldots k_m'\right)
        G_{nm}\left(E,k_1,\ldots k_n, k'_1,\ldots
        k'_m\right),}
\auto
$$
where the $G_{nm}$ are reggeon ``Green's functions'' satisfying a
``unitarity equation'' -
$$
\eqalign{
G_{nm}&\left(E+i\epsilon,k\right)
        -G_{nm}\left(E-i\epsilon,k\right){} =
        {1 \over (2\pi)^{3M}} \sum_M(-1)^{M-1}\int\prod_s d^2k_s\cr
       &\delta^2\left(k-\sum^M_{s=1}k_s\right)
\delta\left(E-\sum^M_{s=1}\Delta(k^2_s)\right)
        G_{nr}\left(E+i\epsilon,k\right) G_{rm}\left(E-i\epsilon,
        k\right).}
\auto
$$
This equation is satisfied by writing Pomeron reggeon
diagrams. For an n-Pomeron state we write the (``non-relativistic") propagator
$$
\eqalign { \Gamma_n = {1
\over \left(E-\sum^n_{i=1} \Delta(k^2_i)\right)}}
\auto
$$
A minimal unitary set of diagrams is obtained by iterating this propagator
with transverse momentum conserving $1\to 2$ and $2\to
1$ triple Pomeron interactions and then integrating over all transverse
momenta. The arbitrariness arising from the
possible existence of an infinite number of unknown higher-order Pomeron
couplings is overcome by demonstrating that (when $\alpha_{\spom}(0)=1$)
there is a fixed point where {\bf only the triple coupling is relevant}. The
result is the well-known Critical Pomeron scaling theory\cite{mpt}.

For an odd-signature Regge pole the corresponding reggeon diagrams are much
more complicated - due to the presence of {\bf vector particle} poles.
It has been known for some time\cite{reg} that the Lipatov equation can be
written as a {\bf reggeon ``Bethe-Salpeter equation''} with the
$O(g^2)$ kernel as a {\bf singular} reggeon interaction. (The singularity of
the kernel is due to gluon poles). Nevertheless, the complete set of reggeon
diagrams should determine (and be determined by) {\bf all Regge limit logs}.
Therefore, if we could construct such diagrams directly we could predict the
results of arbitrary higher-order log calculations.

At first sight, such a construction looks impossible both because of
{\bf singular reggeon interactions due to gluon poles} and {\bf undetermined
parameters in higher-order interactions}. The first problem is resolved in
general by the construction of reggeon loops via multiple discontinuities -
this
explicitly involves {\bf no singular interactions}. The second problem is
resolved for Yang-Mills theories by the {\bf imposition of Ward identity
constraints directly on reggeon amplitudes}. The result is a powerful technique
for constructing higher-order reggeon diagrams, and therefore {\bf higher-order
Lipatov kernels}, without going through the very, very, complicated underlying
Feynman diagram calculations.

We will illustrate the method by outlining a derivation of the
Lipatov equation directly from reggeon diagrams, then present the $O(g^4)$
kernels. We first discuss general properties of the diagrams.

A general reggeon amplitude is gauge-invariant and (in a suitable multi-Regge
limit) can be embedded in a multigluon S-Matrix element as illustrated in
Fig.~1. If $k_{\perp} \to 0$ for one reggeon, then we obtain an amplitude
$A_{\nu}$ for a gluon with zero {\bf four-momentum}
$k_{\mu}$ to couple to a physical multigluon state. This gluon amplitude
satisfies the Ward identity\cite{gth}
$$
\eqalign{ k_{\mu}A_{\mu} = 0}
\auto\label{war}
$$
which, {\bf provided there are no massless fermions in the theory}, implies
$A_{\nu}$ vanishes at $k_{\mu}=0$. Consequently, the reggeon amplitude must
also vanish {\bf when $k_{\perp} = 0$.}

To use multiple discontinuity theory we introduce an $\alpha'$ for
gluons from the outset - we will show that $\alpha' \to 0$ gives
perturbative results. The gauge group is manifest via group factors in
reggeon vertices and gauge invariance is imposed by the Ward identity
constraint we have just discussed. In addition to reggeon propagators,
diagrams will now contain a particle pole or ``signature factor''
$[\alpha'k^2]^{-1}$ for each {\bf uncut} reggeon line.
The triple reggeon vertex contains a {\bf ``nonsense zero''} i.e.
$$
\eqalign{\Gamma_{12}~\sim~ ~g~\sqrt{\alpha'}~[E - \alpha'k_1^2 -
\alpha'k_2^2]~\centerunder{$\sim$} {\raisebox{-6mm} {$(E=\alpha'k^2)$}}~
[k^2 - k_1^2 -k_2^2]~~~
\centerunder{$\sim$} {\raisebox{-6mm} {$(k_1^2=k_2^2=0)$}}~
{}~ k^2 }
\auto
$$

The simplest reggeon diagrams are the set of ``self-energy'' diagrams.
These give gluon reggeization. In the cut diagrams
- {\bf all two reggeon propagators are cancelled by nonsense
zeroes}, leaving one zero per loop to provide the reggeization. The
result is a series which sums up to
$$
\eqalign{[E - \alpha'k^2 - g^2~k^2J_1(k^2)]^{-1}
\centerunder{$\longrightarrow$} {\raisebox{-5mm} {$\alpha' \to 0$}}
[E - g^2~k^2J_1(k^2)]^{-1}}
\auto
$$
where
$$
\eqalign{J_1(k^2)
{}~~=~~{1 \over (2\pi)^3}\int d^2q {1 \over q^2(k-q)^2}}
\auto
$$
which is the {\bf the perturbative reggeization result}.

To construct multi-loop diagrams we have to proceed loop-by-loop,
utilising multiple discontinuities. E.g. for the two-loop diagram
shown in Fig.~2, we first construct the {\bf triple discontinuity} of a
three-reggeon vertex diagram. As illustrated in Fig.~3, {\bf nonsense zeroes at
the vertices cancel the reggeon propagators $\Gamma_2$ and $\Gamma_3$} (and
produce an external zero factor). The cuts also remove an internal signature
factor. {\bf Only a transverse momentum integral remains} which is then used
as a vertex in the one-loop bubble diagram to obtain the two-loop diagram.

The two-loop diagrams in the {\bf color zero}, even-signature, channel are
shown in Fig.~4. Now {\bf only the three reggeon propagators are cancelled
by nonsense zeroes}. With external couplings $\alpha'g^2$, the sum of such
diagrams gives
$$
\eqalign{ {g^4 \over (2\pi)^6} &\int {d^2k_1 \over k_1^2} {d^2k_2 \over
k_2^2} \delta^2(k-k_1-k_2)
\int {d^2k_3 \over k_3^2} {d^2k_4 \over k_4^2} \delta^2(k-k_3-k_4)\cr
&{1 \over (E-\alpha'k_1^2 - \alpha'k_2^2)}
{1 \over (E-\alpha'k_3^2 - \alpha'k_4^2)} ~K^{(2)}_{2,2}(k_1,k_2,k_3,k_4)}
\auto\label{6th}
$$
where $K^{(2)}_{2,2}(k_1,k_2,k_3,k_4)$ is a {\bf sum of transverse momentum
diagrams}. The relative weight of the diagrams is uniquely
determined by demanding both {\bf infra-red finiteness} of the (integral)
kernel and the {\bf Ward identity} vanishing when $k_i \to 0,~i=1,..,4$. We
thus obtain, {\bf without calculating a single Feynman diagram},
the Lipatov kernel, to $O(g^2)$, given in (\ref{2,2}) above.
The limit $\alpha' \to 0$ of (\ref{6th}) gives directly the sixth-order
perturbative result (if we identify $g$ with the gauge coupling). Iteration of
the construction procedure to obtain the full Lipatov equation is
straightforward.

The $O(g^4)$ contribution to the (2-2) Lipatov kernel originates
from three-loop reggeon diagrams of the form shown in Fig.~5 and the resulting
kernel is a sum of the corresponding transverse momentum diagrams.
{\bf If the vanishing at $k_i \to 0,~i=1,..,4$, is imposed together with
infra-red finiteness, the relative weights of the distinct transverse
momentum diagrams is uniquely determined} and the result is
$$
\eqalign{g^{-4} K^{(4)}_{2,2}(k_1&,k_2,k_3,k_4)~=~
\sum_{\scriptscriptstyle 1<->2}~
\Biggl(~{2 \over 3}(2\pi)^3 k_1^2J_2(k_1^2)k_2^2\Bigl(k_3^2\delta^2(k_2-k_4)
+~k_4^2\delta^2(k_2-k_3)\Bigr)~\cr
&-~\Biggl({k_1^2J_1(k_1^2)k_2^2k_3^2~+~k_1^2J_1(k_1^2)k_2^2k_4^2~+~
k_1^2k_3^2J_1(k_3^2)k_4^2~+~k_1^2k_3^2k_4^2J_1(k_4^2) \over
(k_1-k_3)^2} \Biggr)\cr
&~+~J_1((k_1-k_3)^2)\Bigl(k_2^2k_3^2+k_1^2k_4^2\Bigr)
{}~+~k_1^2k_2^2k_3^2k_4^2~I(k_1,k_2,k_3,k_4)~\Biggr) }
\auto
$$
with
$$
\eqalign{ I~=~{1 \over (2\pi)^3}\int d^2q {1 \over
q^2(q+k_1)^2(q-k_3)^2(q+k_1-k_4)^2}~~,~~
J_2~=~{1 \over (2\pi)^3}\int d^2q {1 \over (k-q)^2}J_1(q^2)}
\auto
$$

Similiarly a {\bf new (2-4) kernel} is generated which is also {\bf uniquely
determined by the Ward identity and infra-red finiteness constraints}
$$
\eqalign{& K^{(4)}_{2,4}(k_1,..,k_6)=
\sum_{\scriptscriptstyle 1<->2}
2\pi^3k_2^2\Biggl(\delta^2(k_2-k_6)K^{(4)}_{1,3}(k_1,k_3,k_4,k_5)
+\delta^2(k_2-k_5)K^{(4)}_{1,3}(k_1,k_3,k_4,k_6)\cr
&~+~\delta^2(k_2-k_4)K^{(4)}_{1,3}(k_1,k_3,k_5,k_6)
{}~+~\delta^2(k_2-k_3)K^{(4)}_{1,3}(k_1,k_4,k_5,k_6)\Biggr)
-~K^{(4)}_{2,4}(k_1,..,k_6)_c}
\auto
$$
where $K^{(4)}_{1,3}$ and $K^{(4)}_{2,4}(k_1,..,k_6)_c$ are given by
$$
\eqalign{ K^{(4)}_{1,3}(k,k_3,k_4,k_5)~=~{1 \over (2\pi)^3}\int
{d^2k_1 \over k_1^2}{d^2k_2 \over k_2^2}~k^2\delta^2(k-k_1-k_2)~
K^{(4)}_{2,3}(k_1,k_2,k_3,k_4,k_5)}
\auto
$$
with
$$
\eqalign{& g^{-4}K^{(4)}_{2,3}(k_1,..,k_5)~=~\sum_{\scriptscriptstyle
1<->2}\Biggl(~(k_1+k_2)^2
{}~-~\Biggl( {k_1^2(k_4+k_5)^2 \over (k_1-k_3)^2}
{}~+~{k_1^2(k_3+k_5)^2 \over (k_1-k_4)^2}\cr
&~+~{k_1^2(k_3+k_4)^2 \over (k_1-k_5)^2} \Biggr)
{}~+~{1 \over 3}\Biggl( {k_1^2k_5^2 \over (k_2-k_5)^2}
{}~+~{k_1^2k_4^2 \over (k_2-k_4)^2}
{}~+~{k_1^2k_3^2 \over (k_2-k_3)^2} \Biggr)\cr
&~+~{2 \over 3}\Biggl( {k_1^2k_2^2k_4^2 \over (k_1-k_3)^2(k_2-k_5)^2}
{}~+~{k_1^2k_2^2k_5^2 \over (k_1-k_4)^2(k_2-k_3)^2}
{}~+~{k_1^2k_2^2k_3^2 \over (k_1-k_5)^2(k_2-k_4)^2} \Biggr)\Biggr)}
\auto
$$
and
$$
\eqalign {& g^{-4}K^{(4)}_{2,4}(k_1,k_2,k_3,k_4,k_5,k_6)_c~=~
\sum_{\scriptscriptstyle 1<->2}~\Biggl(~(k_1 + k_2)^2
- \Biggl( {k_1^2(k_4+k_5+k_6)^2 \over (k_1-k_3)^2}\cr
&+{k_1^2(k_3+k_5+k_6)^2 \over (k_1-k_4)^2}
+{k_1^2(k_3+k_4+k_6)^2 \over (k_1-k_5)^2}
+{k_1^2(k_3+k_4+k_5)^2 \over (k_1-k_6)^2}\Biggr)
-{1 \over 4}\Biggl( {k_1^2k_3^2 \over (k_2-k_3)^2}\cr
&+{k_1^2k_4^2 \over (k_2-k_4)^2}
+{k_1^2k_5^2 \over (k_2-k_5)^2}
+{k_1^2k_6^2 \over (k_2-k_6)^2}\Biggr)
+{1 \over 2}\Biggl( {k_1^2(k_5+k_6)^2 \over (k_2-k_5-k_6)^2}
+{k_1^2(k_5+k_4)^2 \over (k_2-k_5-k_4)^2}\cr
&+{k_1^2(k_4+k_6)^2 \over (k_2-k_4-k_6)^2}
+{k_1^2(k_3+k_6)^2 \over (k_2-k_3-k_6)^2}
+{k_1^2(k_5+k_3)^2 \over (k_2-k_5-k_3)^2}
+{k_1^2(k_3+k_4)^2 \over (k_2-k_3-k_4)^2}\Biggr)\cr
&+{1 \over 2}\Biggl( {k_1^2k_2^2(k_4+k_5)^2 \over (k_1-k_3)^2(k_2-k_6)^2}
+{k_1^2k_2^2(k_3+k_5)^2 \over (k_1-k_4)^2(k_2-k_6)^2}
+{k_1^2k_2^2(k_3+k_4)^2 \over (k_1-k_5)^2(k_2-k_6)^2}\cr
&+{k_1^2k_2^2(k_3+k_6)^2 \over (k_1-k_4)^2(k_2-k_5)^2}
+{k_1^2k_2^2(k_4+k_6)^2 \over (k_1-k_3)^2(k_2-k_5)^2}
+{k_1^2k_2^2(k_5+k_6)^2 \over (k_1-k_3)^2(k_2-k_4)^2}\Biggr)\cr
&-{1 \over 4}\Biggl( {k_1^2k_2^2k_4^2 \over (k_1-k_3)^2(k_2-k_5-k_6)^2}
+{k_1^2k_2^2k_5^2 \over (k_1-k_3)^2(k_2-k_4-k_6)^2}
+{k_1^2k_2^2k_6^2 \over (k_1-k_3)^2(k_2-k_4-k_5)^2}\cr
&+{k_1^2k_2^2k_3^2 \over (k_1-k_4)^2(k_2-k_5-k_6)^2}
+{k_1^2k_2^2k_5^2 \over (k_1-k_4)^2(k_2-k_3-k_6)^2}
+{k_1^2k_2^2k_6^2 \over (k_1-k_4)^2(k_2-k_3-k_5)^2}\cr
&+{k_1^2k_2^2k_4^2 \over (k_1-k_5)^2(k_2-k_3-k_6)^2}
+{k_1^2k_2^2k_3^2 \over (k_1-k_5)^2(k_2-k_4-k_6)^2}
+{k_1^2k_2^2k_6^2 \over (k_1-k_5)^2(k_2-k_4-k_3)^2}\cr
&+{k_1^2k_2^2k_4^2 \over (k_1-k_6)^2(k_2-k_5-k_3)^2}
+{k_1^2k_2^2k_5^2 \over (k_1-k_6)^2(k_2-k_4-k_3)^2}
+{k_1^2k_2^2k_3^2 \over (k_1-k_6)^2(k_2-k_4-k_5)^2} \Biggr)\Biggr)}
\auto\label{k424c}
$$

This technique explicitly combines the defining Ward
identities of the theory with unitarity in the Regge limit. Since this should
be sufficient to define the theory it appears that {\bf Yang-Mills reggeon
theories can be constructed directly}. That is, the Feynman diagram expansion
can be bypassed completely. It certainly looks {\bf straightforward to
construct even higher-order kernels}. There are, of course, many issues to
be studied already with the $O(g^4)$ kernels, including the new Pomeron
intercept, consequences for the anomalous dimensions of two and four gluon
operators, consequences for deep-inelastic high-mass diffraction
etc.\cite{bar}.

\subhead{3. Confinement and Quarks}

The two defining properties we have used for pure glue reggeon theories
are actually sufficient to show that such a theory {\bf can not produce a
confining Pomeron RFT}. While there is an {\bf extensive exponentiation} of
infra-red divergences in color non-zero channels, which removes all color
non-zero amplitudes from the theory,  {\bf all color zero reggeon amplitudes
are infra-red finite}. Also, the Ward identity constraint that reggeon
amplitudes vanish when any $k_{\perp} \to 0$ guarantees that {\bf
phase-space integrals for multi-reggeon intermediate states are finite.}
Consequently {\bf color zero} multigluon {\bf reggeon states }
necessarily survive and give singularities (but not divergences) in all
amplitudes at zero transverse momentum. This clearly implies there is
{\bf no confinement}.

As we now discuss, this conclusion is avoided, {\bf when quarks are
present},
in a manner which does lead to confinement. To argue that the Ward Identity
(\ref{war}) requires that $A_{\nu}$ vanishes at $q=0$ we simply differentiate
$$
\eqalign{ A_{\nu}~+~{\partial A_{\mu} \over \partial q_{\nu}}~q_{\mu}~=0~
=>~A_{\nu}~\centerunder{$\to$} {\raisebox{-5mm} {$(q_{\mu}\to
0)$}}~~0~~unless~~{\partial A_{\mu} \over \partial q_{\nu}}
{}~\to~\infty}
\auto\label{inf}
$$
The Ward identity follows directly from gauge invariance and
can not be violated, {\bf but massless quark infra-red
divergences} of $A_{\nu}$ - arising from the Regge limit - can prevent the
conclusion that $A_{\nu}$ vanishes at $q~=~0$.
There are {\bf quark reggeon diagrams} for which this is the case.
In the color zero two gluon channel, quark loops of the form
shown in Fig.~6 satisfy the Ward Identity
by ``$A_{\mu}~\equiv~A_{-}$'' and ``$q_{\mu}~\equiv~q_{\perp}$'', but
produce a vertex (for quark mass m)
$$
V_c(q,k,k^\prime) \sim \int d^2p\
{{Tr\left[ (-\st{p}+\st{q}+m)\st{k} (-\st{p}-\st{q} +m)\st{k}^\prime \right]}
\over {\left[(p+q)^2 +m^2\right]\left[(p-q)^2 +m^2\right]}}
$$
which satisfies
$$
\eqalign{
V_c ~\centerunder{$\longrightarrow$}{\centerunder
{$m\sim 0$} {\raisebox{-4mm} {$q\sim 0$}}}~~
8 \pi \left[ {{2(q\cdot k)(q\cdot k^\prime)} \over {q^2}}
- k \cdot k^\prime\right]
{}~~\centerunder{$\longrightarrow$}
{\raisebox{-4mm} {$q\rightarrow \pm k, \pm k^\prime$}} ~
8\pi k\cdot k^\prime }
\auto
$$
and so the {\bf Ward identity constraint is not satisfied.} ( This is
consistent with (\ref{inf}) because $~[{\partial V_c
\over \partial q_0}]_{q_0\sim 0} ~~(\equiv [{\partial V_c
\over \partial m}]_{m\sim 0})~~ \sim {1 \over q_0}~~\to \infty$ due to quark
propagator infra-red divergences).

$V_c$ (as a partial-wave amplitude) is distinct from the
Lipatov kernel, but ``~$V_c^2$~'' does contribute to this kernel.
Consequently {\bf a new infra-red
divergence} is generated which produces the {\bf exponentation to zero} of
the two-gluon reggeon state. In fact, in SU(2) gauge theory all color zero
reggeon states produce divergences which are similarly
exponentiated\cite{arw2}, apart from a scaling divergence associated with the
``anomalous Odderon" three gluon state. This divergence couples only via a
``{\bf triangle
quark loop anomaly''} and so does not exponentiate. Instead it produces a
{\bf reggeon} (``winding number'') {\bf condensate} which gives
{\bf a confinement spectrum}. In {\bf SU(3)} gauge theory, a reggeized gluon
in the background of this condensate gives {\bf a Regge pole Pomeron}.

\subhead{4. Pomeron Dynamics and Electroweak Symmetry Breaking}

A Regge pole Pomeron gives $\sigma_T \to 0$ ~{\bf unless ~
$\alpha_{\spom}(0)=1$}. Therefore, (confining) Pomeron RFT is,
asymptotically, inconsistent with the validity of perturbative QCD at large
$k_{\perp}$ {\bf unless the Pomeron is Critical~!!}
My detailed analysis\cite{arw2} of QCD Pomeron RFT suggests
that $\alpha_{\spom}(0)=1$ if, and only if, the number of flavors $N_f$
is a maximum. This requires that $N_f=16$
{\bf or $ N_f=6$ for color triplet quarks and $N_f=2$
for color sextet quarks}.
It is a remarkable coincidence that two flavors of color sextet quarks
can provide {\bf a natural form of dynamical symmetry-breaking\cite{wjm} for
the
electroweak interaction which meshes
perfectly with the observed experimental features.}

Add a massless doublet $(u_6,d_6)$ with the {\bf usual quark quantum
numbers} to the Standard Model (with no scalar Higgs sector).
The axial $U(2)\otimes U(2)$ chiral symmetry breaks spontaneously, producing
four pseudoscalar Goldstone bosons ($\pi^+_6,\;\pi^-_6,\;\pi^0_6$ and
$\eta_6$) which couple ``longitudinally'' to the sextet
axial currents i.e.
$$
\eqalign{ <0|A^\tau_{\mu}|\pi^{\tau}_6(q)>~\sim F_{\pi_6}q_{\mu}~~,~~
<0|a_{\mu}|\eta_6(q)>~\sim F_{\eta_6}q_{\mu}}
\auto
$$
The $SU(2)$ gauge fields $B^{\tau}_{\mu}$ couple via
$$
\eqalign{{\cal L}_I=gW^{\tau\mu}\Bigl(V^{\tau}_{\mu}-A^{\tau}_{\mu}\Bigr)
}\auto
$$
As a result $\pi^+_6,\pi^-_6$ and $\pi^0_6$ are
``eaten'' and provide {\bf masses for the $W^+,\;W^-$ and $Z^0$}, with
$$
\eqalign{M_W\sim g\;F_{\pi_6}}
\auto
$$
where {\bf $F_{\pi_6}$ is a
QCD scale}. The ``Casimir Scaling'' rule\cite{wjm} gives
$$
\eqalign{C_6~\alpha_s(F^2_{\pi_6})~\sim~C_3~\alpha_s(F^2_{\pi})}
\auto
$$
which is consistent with $F_{\pi_6}\sim 250$ GeV ! Consequently, {\bf the
electroweak scale is naturally explained as a second $QCD$ scale.}
The restriction to a sextet flavor doublet, necessarily gives
$$
\eqalign{\rho=~(M^2_W/M^2_Zcos^2\theta_W)~=~1}
\auto
$$
as also required by experiment.

Finally we note that the sextet sector may also be deeply tied to the issue
of Strong $CP$ conservation and the origin of $CP$ violation.
The $\eta_6$ is a Goldstone boson associated with the $U(1)$
axial chiral symmetry and may be {\bf the axion}. It is a conventional
(Peccei-Quinn) axion\cite{pq} except that it can aquire an {electroweak
scale mass} as a result of enhanced electroweak scale color instanton
interactions\cite{arw3}.

Not only may the sextet sector be the explanation of {\bf Strong $CP$
conservation in the triplet sector} (via the $\eta_6$ axion), but it may
also be responsible for {\bf $CP$ violation at the weak scale}. Because the
sextet sector has no axion the QCD interactions at this scale will naturally
be ``Strong $CP$-violating''. The normal triplet quark hadrons will contain
a {\bf small admixture of sextet quark states} which could {\bf provide
their $CP$ violating interactions}.

\subhead{5. Conclusion}

{\bf The QCD Pomeron may be the key to many of the
remaining puzzles of the Standard Model.}

\newpage

{}~
\vspace{2.5in}

\noindent Fig.~1 A reggeon amplitude gives a gluon amplitude as $k_{\perp}
\to 0$.

\vspace{2in}

\noindent Fig.~2 A two-loop reggeon diagram.

\vspace{2.5in}

\noindent Fig.~3 The reduction of a three-reggeon vertex diagram to a
transverse momentum diagram.

\newpage

{}~
\vspace{2in}

\noindent Fig.~4 Two-loop even-signature reggeon diagrams.

\vspace{3in}

\noindent Fig.~5 Three loop diagrams giving the $O(g^4)$ kernel.

\vspace{2in}

\noindent Fig.~6 A one loop quark reggeon diagram.

\end{document}